# Discrete-element model for the interaction between ocean waves and sea ice

Zhijie Xu[1,a], Alexandre M. Tartakovsky[1] and Wenxiao Pan[1]

1. Computational Mathematics Group, Fundamental and Computational Sciences Directorate, Pacific Northwest National Laboratory, Richland, WA 99352, USA

**Abstract**

We present a discrete element method (DEM) model to simulate the mechanical behavior of sea ice in response to ocean waves. The interaction of ocean waves and sea ice can potentially lead to the fracture and fragmentation of sea ice depending on the wave amplitude and period. The fracture behavior of sea ice is explicitly modeled by a DEM method, where sea ice is modeled by densely packed spherical particles with finite size. These particles are bonded together at their contact points through mechanical bonds that can sustain both tensile and compressive forces and moments. Fracturing can be naturally represented by the sequential breaking of mechanical bonds. For a given amplitude and period of incident ocean wave, the model provides information for the spatial distribution and time evolution of stress and micro-fractures and the fragment size distribution. We demonstrate that the fraction of broken bonds, $\alpha$, increases with increasing wave amplitude. In contrast, the ice fragment size $l$ decreases with increasing amplitude. This information is important for the understanding of breakup of individual ice floes and floe fragment size.

---
a) Electronic mail: zhijie.xu@pnnl.gov Tel: 509-372-4885

**I. Introduction**

The purpose of this paper is to present a discrete element model (DEM) for understanding ocean wave/sea ice interactions and the mechanical response of sea ice to waves with various amplitudes and periods. This information is particularly important in the study of ice margin dynamics in the Marginal Ice Zone (MIZ), for it helps in the understanding of how ocean wave/sea ice interactions are related to the breakup of individual ice floes and in determining the floe size distribution in the entire MIZ [1, 2].

Sea ice may fill inlets and harbors of the Arctic Ocean and the Antarctic Continent [1]. Most ice fields are shielded from direct interaction with the outer ocean waves and may grow over many years. The boundary zone between the open and ice-covered sea is referred to as the MIZ, which consists of many individual ice floes with various shapes and types. Ocean waves play an important role in ice dynamics in the MIZ because they are the primary energy source that is responsible for the breakup or fragmentation of sea ice [1].

The effect of ocean waves in ice dynamics is well documented in the literature [3-5]. In principle, wave energy propagates in the form of flexural-gravity waves in ice floes accompanied by the energy loss due to the wave scattering at imperfections, the ice creep deformation, and the floe collision that leads to the wave attenuation. The ice floe could be significantly deformed while the flexural-gravity wave penetrates into it. Depending on the magnitude and the frequency of the ocean wave, the fracturing can occur if the stress or strain induced in the ice is greater than the ice can sustain. This provides an important mechanism for the breakup of a vast ice field into many pieces of floes. The thermodynamics of ice grow and ice melt can be significantly changed due to the breakup of ice floes where the ice melting is accelerated during the summertime and the ice formation is enhanced during the wintertime [6, 7].

The mechanical behavior of ice is significantly affected by the nucleation and growth of micro-fractures. Realistic and robust models for sea ice should account for these details at small and large scales. The Discrete Element Method described in this paper is an approach to approximate complex materials as assemblies of independent discrete elements (particles) of various sizes, shapes, and other properties that interact via cohesive interactions, repulsive forces, and friction forces. The macroscopic behavior can be treated as a collective behavior of many interacting discrete elements.

DEM was first introduced by Cundall [8-10] as an alternative to continuum mechanics. It has been extensively applied to simulations of ball mills [11], the shear flow of non-cohesive granular materials [12], the behavior of crushable soil agglomerates [13], and the mechanical behavior of rocks [14]. Recently, Wilchinsky etc. introduces a DEM model to the study of the effect of wind stresses on the rupture behavior of ice pack and pattern of faults due to the ice mechanical failure [15, 16]. In DEM, both the grains and mechanical bonds are deformable, and the bonds may break when either the tensile or shear stress exceeds the critical strength. Similar to other popular particle methods such as Smoothed Particle Hydrodynamics (SPH) [17, 18] and Dissipative Particle Dynamics (DPD) [18-20], the movement of each DEM particle, including translation and rotation, can be calculated from Newton's second law through explicit numerical integration in the fashion of molecular dynamics (MD).

In this paper, we describe a DEM model for modeling the mechanical behavior of sea ice due to the interaction with ocean waves of various amplitudes and periods. The fracturing behavior is expected to be important in this application and is included naturally in the DEM model. The model provides important insights for a better understanding of the breakup of sea ice in the MIZ. This paper is organized as follows. Section II describes the general Discrete Element Method, followed by the parameterization of a DEM model for sea ice in Section III and a DEM model for ocean wave/sea ice interaction in Section IV. Section V provides the numerical results and discussion.

## II. Discrete Element Method

In the standard DEM, the computational domain is discretized into a collection of circular (2D) or spherical (3D) particles of various sizes. Each particle has a finite size and mass, and the particles are kept together by bonds at their points of contact. The contact forces (normal and shear force) and moments (in-plane twist and out-of-plane bending moment) between DEM particles can be calculated in an incremental fashion or from their relative displacements. The displacement of each particle including translation and rotation can be calculated from Newton's second law through an explicit numerical integration in the fashion of molecular dynamics (MD). Dynamic fracturing can be naturally modeled through the sequential breakup of inter-particle bonds.

As shown in Fig.1, particles have a finite stiffness so that two particles are allowed to overlap in a small region relative to their size. The overlap between the particles results in a contact force. The bond between two particles is treated as a mechanical element that also has a finite stiffness and can carry both force and moment. In Fig 1, the bond (gray area), connecting particles *A* and *B*, can break depending on the interactions between two particles. In summary, particles *A* and *B* interact with each other through both contact force and the mechanical bond. The total resultant force and moment due to the interaction between two particles are therefore comprised of the contact contribution ($F_g$) and the bond contribution ($F_b$ and $M_b$), namely the contribution from the grain-grain interaction and from the mechanical bond, respectively.

The contact force contribution $F_g$ can be resolved into the normal ($F_g^n$) and shear components ($F_g^s$) with respect to the contact plane between particles *A* and *B*. A linear force-displacement law is employed for each component, where

$$F_g^n = K^n u^n, \tag{1}$$

and

$$\Delta F_{g,t}^s = -K^s \Delta u_{,t}^s. \tag{2}$$

$K^n$ and $K^s$ are the grain normal and shear stiffness, and $u^n$ is the displacement along the centerline connecting *A* and *B*. Shear component $F_{g,t}^s$ is computed in an incremental fashion (Eq. (2)) in the sense that $\Delta F_{g,t}^s$, the increment of shear component $F_g^s$ at time *t*, is computed at each time step based on the shear displacement increment $\Delta u_{,t}^s$ (perpendicular to the centerline). Therefore, the total shear component at the next time step can be written as,

$$F_{g,t+\Delta t}^s = F_{g,t}^s + \Delta F_{g,t}^s. \tag{3}$$

The bond contribution, $F_b$, to the total force can be resolved into normal ($F_b^n$) and shear components ($F_b^s$) in a similar fashion:

$$F_b^n = k^n A u^n, \tag{4}$$

and

$$\Delta F_{b,t}^{s} = -k^{s} A \Delta u_{,t}^{s}. \tag{5}$$

Here, $k^n$ and $k^s$ are the bond normal and shear stiffness per unit area. $A$ is the area of the bond cross-section and chosen to be $A = \pi R^2$, where $R = \min(R^{(A)}, R^{(B)})$ is radius of the bond chosen to be the smaller size of particles $A$ and $B$. The total shear component $F_{b,t}^{s}$ can be computed as,

$$F_{b,t+\Delta t}^{s} = F_{b,t}^{s} + \Delta F_{b,t}^{s}. \tag{6}$$

The bond moment contribution $M_b$ can be resolved into the bending moment ($M_b^s$, acting out of contact plane) and the twisting moment ($M_b^n$, acting in contact plane). Linear relationships are established between the moments and the bending and twisting angles,

$$M_b^s = -k^s I \theta^s, \tag{7}$$

and

$$M_b^n = -k^n J \theta^n, \tag{8}$$

where $I$ and $J$ are the moment of inertia and polar moment of inertia, respectively. $\theta^s$ and $\theta^n$ are the bending and twisting angles. The maximum normal ($\sigma_b$) and shear stress ($\tau_b$) in the cross-section of mechanical bond can be easily calculated with given contact and bond contributions,

$$\sigma_b = -\frac{F_b^n}{A} + \frac{|M_b^s| R}{I}, \tag{9}$$

and

$$\tau_b = \frac{|F_b^s|}{A} + \frac{|M_b^n| R}{J}, \tag{10}$$

where $R$ is the radius of the bond cross-section. The mechanical bond can break if either normal ($\sigma_b$) or shear stress ($\tau_b$) in the bond is beyond its strength (threshold value), namely $\sigma_b > \sigma_b^{MAX}$ or $\tau_b > \tau_b^{MAX}$. There still exists grain-grain interaction (contact force) if two particles come back into contact even after bond is broken. In principle, bond strengths $\sigma_b^{MAX}$ and $\tau_b^{MAX}$ can be various numbers for different bonds to mimic the effect of random defects. In the current study, we are using same strengths for every bond. At any time step, all bonds with stresses large than threshold values are removed from the model along with all associated forces and moments. By this manner, the complicated micro-fractures and fracture network can be represented by the broken bonds, and the fracturing process can be modeled as the sequential breakup of mechanical bonds.

## III. Parameterization of the DEM Sea Ice Model

The DEM model parameters can be calculated using the mechanical properties of sea ice. We use the following material constants for wave/ice interactions and ice mechanical properties: the Young's modulus of ice $E = 6 \ GPa$, the Poisson ratio $\nu = 0.3$, the density of water $\rho_w = 1025 \ (kg/m^3)$, and the density of ice $\rho_i = 922.5 \ (kg/m^3)$ [1, 21]. The ice tensile and compressive strength have been measured by a number of researchers [22]. In general, the ice strengths are dependent on temperature, strain rate, and many other factors. The tensile strength $\sigma_t$ has a wide range from 0.7 to 3.1 MPa, with an average strength of 1.43 MPa for the temperature range of -10ºC to -20ºC. The compressive strength $\sigma_c$ ranges from 5 to 25 MPa in the same temperature range [22].

An example of the relationship between the material modulus and the DEM parameters is provided in [14]. Input parameters in the DEM sea ice model should be chosen to match the ice strength. We implement simulations for uniaxial tests to calibrate DEM model parameters with ice mechanical properties. Inset of Fig.2 shows the geometry used in uniaxial tests with a width of $w$=2.5m and a height of $h$=5m. Periodic boundary conditions were applied to the vertical boundaries, and both upper and lower horizontal boundaries subject to a constant vertical speed to mimic the constant strain rate tests. A time step $\Delta t = 1.24 \times 10^{-5} s$ was chosen for explicit integration so that any disturbance cannot propagate farther than its neighbors in a single step [14]. Figs. 2 and 3 show the macroscopic stress-strain curves for uniaxial tensile and compressive tests obtained directly by DEM simulations with parameters:

$$K^n = ED \text{ and } K^s = 0.4ED, \tag{11}$$

$$k^n = E/D \text{ and } k^s = 0.4\,E/D, \tag{12}$$

$$\sigma_b^{MAX} = 0.001E \text{ and } \tau_b^{MAX} = 0.001E, \tag{13}$$

where $D = 0.1m$ is the average diameter of DEM particles. As adopted in most DEM simulations, the size of the DEM particles is assumed to follow a uniform distribution in the range [0.075m, 0.125m] with an average size of $D = 0.1m$. A Poisson ratio of $v$=0.25 is used for the ice material and the prefactor 0.4 in Eqs. (11) and (12) is a direct result of the relationship between shear and Young's modulus for isotropic materials, where $K^s = K^n/2(1+v)$. The corresponding tensile and compressive strength of the ice obtained from the DEM simulations are $\sigma_t = 2.7 MPa$ and $\sigma_c = 9.7 MPa$, well within the range available from the literature [22]. In general, both the size distribution and the ice strength can have effects on the ice fragmentation behavior, which can be carefully investigated by the same simulation methodology but with different DEM parameters.

In reality, the microstructure of natural sea ice largely depends on the ice formation processes. For example, the granular ice forms under dynamic and turbulent conditions while the columnar ice forms under static conditions. The real sea ice is a complex heterogeneous and anisotropic material depending on the forming process. In the current study, sea ice was treated as a homogenous, isotropic material in the current model. However, heterogeneity can be naturally introduced into the model through the position-dependent DEM parameters, where the effect of heterogeneity on ice fragmentation can be studied in more detail.

**IV. DEM Model for Ocean Wave/Sea Ice Interactions**

The set of parameters described in Section III was used in simulations of wave/ice interactions, where an ice field of length $L$ and a uniform thickness $h=1m$ is assumed to float over ocean water of depth $H=100m$. We considered the wave/ice interactions through the reflection and transmission of waves at the edge of sea ice [1, 21]. It has been observed that both reflection and transmission are dependent on the period $\Omega$ of the incident wave, the thickness $h$ of sea ice, the ice mechanical properties including the Young's modulus $E$ and Poisson ratio $v$, the sea water density $\rho_w$ and ice density $\rho_i$, and the water depth $H$ over which the ice is covering [1, 21, 23]. Intuitively, a thinner ice permits a greater transmission, and a thicker ice leads to a larger reflection at the ice edge. Reflection and transmission coefficients $R$ and $T$ can be defined in terms of the amplitudes of surface displacement for incoming, reflecting, and transmitting waves,

$$R = B_1/A_1 \text{ and } T = A_2/A_1, \tag{14}$$

where $A_1$ is the amplitude of the incoming displacement wave $\phi_i = A_1 e^{i(k_1 x - \omega t)}$. The reflected wave is given by $\phi_r = B_1 e^{i(k_1 x - \omega t)}$, and the transmitted wave is defined as $\phi_t = A_2 e^{i(k_2 x - \omega t)}$, with $B_1$ and $A_2$ being the amplitudes. $k_1$ and $k_2$ are wave numbers of incident and transmitted waves.

Fox and Squire [21] proposed a mathematical model for the reflection and transmission coefficients at the edge of sea ice, where the sea ice is modeled as a continuous, thin and elastic plate of uniform thickness. We use the Bernoulli-Euler theory for an elastic thin plate to describe the flexural-gravity wave propagation in sea ice. Through the matching of solutions at the interface between open water and the water covered with sea ice, the model is able to compute $R$ and $T$ with any given wave period $\Omega$. Figure 4 shows the variation of coefficients $R$ and $T$ with period $\Omega$ of the incident wave, a reprint from Fox and Squire's work in [21].

We use the DEM model to study the interaction between ocean waves and sea ice. Specifically, we study the effect of amplitude and period of an ocean wave on the fracturing of sea ice. In general, an ocean wave with larger amplitude $A_2$ and smaller wavenumber $k_2$ has a greater potential to break the ice floe than a wave with smaller amplitude and larger wavenumber. Furthermore, incident waves with shorter periods result in transmitted waves with smaller transmitted coefficients $T$ but larger wavenumber $k_2$.

In the DEM simulations, a two-dimensional ice floe was generated as shown in Fig. 5. The amplitude $A_2$ and the wavenumber $k_2$ for transmitted wave $\phi_t$ are calculated from the incident wave $\phi_i$ according to Fox and Squire [21], and are used to apply load in the DEM model. Implementation of boundary conditions on complex stationary or moving boundaries is challenging for particle methods [20]. The phase-field approach [24, 25] is based on the concept of diffuse interface and can be used to provide an accurate way to represent boundaries for particle methods. In this study, difficulties associated with the boundaries are removed as we do not explicitly model the interaction between water and ice. In order to generate a displacement field with the form of $u(x,t) = A_2 \cos(k_2 x - \omega t)$ (the real part of $\phi_t$), the ice floe was first slowly deformed to an initial deformation of $u(x) = A_2 \cos(k_2 x)$. A prescribed velocity $v(x,t) = A_2 \omega \sin(k_2 x - \omega t)$ in y direction was applied to all DEM particles to deform the ice floe and generate the desired displacement field $u(x,t)$.

In summary, to simulate deformation of the ice due to wave/ice interaction, a vertical velocity (in y direction) $v(x,t) = A_2 \omega \sin(k_2 x - \omega t)$ is applied to all DEM particles forming the ice floe. The amplitude $A_2$ and the wavenumber $k_2$ for transmitted wave $\phi_t$ with given period $\Omega$ (or angular frequency $\omega$) are found from the amplitude $A$, the wavenumber $k_1$ and the period $\Omega$ of the incident wave.

We perform two sets of numerical experiments to simulate the response of the sea ice subjected to incident waves with different wave periods $\Omega$ and amplitudes $A_1$. Each set includes 7 simulations with various wave amplitudes. Table 1 summarizes parameters used in all simulations. In the first set of experiments, the sea ice is subjected to an incident wave with longer period than the second set.

## V. Results and Discussion

As shown in Fig. 5, a total number of ~10,000 DEM particles with an average size of ($D$=0.1m) are used to model the ice with a uniform thickness of $h = 1m$ and a length of $L = 2\pi/k_2$. The length $L$ is chosen so that at least one entire wavelength of transmitted wave can be investigated.

A periodic boundary condition is applied in $x$ direction and a free boundary is used for the top and bottom surface in all numerical simulations. The traveling transmitted wave is generated in the sea ice structure by prescribing velocity field $v_i(t) = A_2 \omega \sin(k_2 x_i - \omega t)$, where $v_i$ is the vertical velocity of particle $i$ and $x_i$ is $x$-component of the vector position of particle $i$. This leads to inhomogeneous deformation and stress fields in the sea ice. The mechanical bonds between DEM particles are allowed to break where the local deformation produces stress exceeding the critical stress of the ice. The passing transmitted wave gives rise to a complicated and inhomogeneous stress distribution because the dynamic bond-breaking process can redistribute the stress field (stress is relaxed near to the broken bonds). Redistribution of the stress field, in turn, affects the bond breaking (the micro-fracture distribution) in the ice structure. The sea ice can eventually break apart depending on the amplitude and period of the incident ocean wave.

We first present the time evolution of the number of broken bonds that is used to quantitatively describe the extent of damage in the sea ice. In all four cases, the fraction of broken bonds $\alpha$ is monitored against the normalized simulation time $\tau$. The parameters $\alpha$ and $\tau$ are defined as,

$$\alpha = N_{broken}/N_{total}, \tag{15}$$

$$\tau = 2t/\Omega, \tag{16}$$

where $\Omega$ is the period of the transmitted wave. $N_{broken}$ and $N_{total}$ denote the number of broken bonds and total bonds in the DEM model. Figures 6 and 7 show the variation of $\alpha$ with dimensionless simulation time $\tau$ for the simulation sets 1 and 2, respectively. As expected, $\alpha$ (the fraction of broken bonds) is monotonically increasing to an asymptotic value, which is reached at $\tau = 2$ when the transmitted wave travels by one wave period $\Omega$ in all simulations. The final value $\alpha_{max}$ is significantly different for each case depending on the incident wave period and amplitude. Figure 8a) shows the variation of $\alpha_{max}$ with input wave amplitudes and periods. In general, there is only a negligible damage caused in sea ice by incoming waves with small amplitudes (1.125m for Case 1-6, 0.75m for Case 1-7, 2.25m for Case 2-6, and 1.5m for Case 2-7). A significant increase in damage will be observed for waves with larger amplitudes (1.5m for Case 1-5 and 3.0m for Case 2-5). After the initial incubation stage, the damage parameter $\alpha_{max}$ is increasing almost linearly with the incoming wave amplitude. For waves with largest amplitudes (3.0m for Case 1-1 and 6.0m for Case 2-1), $\alpha$ reaches almost 8%. It was also shown that waves with same amplitudes but longer periods will cause much more damage than waves with shorter periods.

If the sea ice is assumed to break apart into fragments with a uniform length $l$ (the assumption is valid for a traveling transmitted wave that is periodic in both space and time), then the relationship $l = 10L/(\alpha N_{total})$ can be obtained, where 10 is the average broken bonds needed to break the ice floe in thickness direction. It is clear that average fragment size ($l$) is inversely proportional to $\alpha$. Figure 8b) shows the dependence of average fragment size $l$ on the incoming wave amplitudes and periods. In general, $l$ is decreasing with increasing amplitude. A minimum fragment size $l \approx 0.7m$ is obtained with the largest amplitudes for both sets of simulations.

In order to examine the overall response of the sea ice to transmitted waves, the deformation patterns for Case 1-1 and Case 2-5 are shown in Fig. 9 and Fig. 10. The snapshots were taken at times $\tau = 0.5$, 1.0, 1.5 and 2.0. The color scale (online only) represents the distribution of stress ($\sigma_{xx}$, normalized by the ice Modulus *E*) with blue (positive) indicating the region under tension and red (negative) the region under compression. In both figures, the generated displacement wave is progressively propagating from the left to the right side. The induced stress wave is also continuously propagating through the entire sea ice structure, and subjecting the ice to repeated bending mode (tensile-compressive-tensile-compressive….). A similar fatigue process was observed in [2]. Fig. 11 shows the stress and fracture distribution, resulting from the repeated bending, for Case 1-1 at $\tau = 2.5$ in the entire ice floe. Fig. 11 also shows zoomed-in pictures around the location with maximum stress. The micro-fractures (broken bonds), denoted by small black arrows, are almost evenly distributed throughout the entire ice floe. The bottom picture in Fig. 11 clearly indicates that macroscopic cracks are almost always in the vertical direction due to the tensile failure. The top picture in Fig. 11 also shows the localized stress due to the generation of micro-fractures in more detail, where the stress is relaxed around the micro-fractures.

The distributions of micro-fractures and the corresponding stress distributions for Cases 1-5, Case 2-5, and Case 2-7 are also shown in Figs. 12-14. As expected, the micro-fractures generated in Cases 1-5 and 2-5 are much fewer, and the average distances between macro-cracks are larger than that in Case 1-1. As a result, larger ice fragments can be formed in Cases 1-5 and 2-5 relative to the Case 1-1. Because of the combined effects of the small amplitude and the large wavenumber, no micro-fractures or broken bonds are observed in Case 2-7.

## VI. Conclusion

A discrete element method (DEM) model was used to simulate the mechanical behavior of sea ice subjected to a passing ocean wave. In the DEM model, an ice floe is represented by densely packed circular particles. To simulate the deformation of sea ice floe due to ocean wave/ice interaction, a velocity field $v(x,t) = A_2 \omega \sin(k_2 x - \omega t)$ is applied to each DEM particle forming the ice floe. The amplitude $A_2$ and the wavenumber $k_2$ are found from the amplitude $A_1$, the wavenumber $k_1$, and the period of the incident wave. The fracturing of sea ice was modeled by computing stresses in bonds connecting adjacent particles. When stresses exceed critical values, bonds are removed and fractures are formed. We demonstrated that the fraction of broken bonds, $\alpha$, increases with increasing amplitude. In contrast, the ice fragment size $l$ decreases with increasing amplitude. The expected fragment size $l$ ($\propto \alpha$) is shown to be highly dependent on the incoming wave period and the amplitude. For example, an increase of amplitude from 1.5m to 3.0m leads to a 75% decrease in size $l$. As an attempt to apply the DEM model to wave/ice interaction, our results show that the DEM model can be used to quantitatively investigate the interactions between sea ice and ocean waves.

**ACKNOWLEDGMENTS**

This research was supported by the Scientific Discovery through Advanced Computing Program of the Office of Science, U.S. Department of Energy. The Pacific Northwest National Laboratory is operated by Battelle for the U.S. Department of Energy under Contract DE-AC06-76RL01830.

Table.1. List of parameters for eight simulation scenarios.

| | Length $L$ (m) | Incident wave $\phi_i$ | | | Transmitted wave $\phi_t$ | |
|---|---|---|---|---|---|---|
| | | Period $\Omega$ (s) | Frequency $\omega$ (1/s) | Amplitude $A_1$ (m) | Wavenumber $k_2$ (1/m) | Amplitude $A_2$ (m) |
| Case1-1 | 104.7 | 6.28 | 1 | 3.00 | 0.06 | 1.00 |
| Case 1-2 | | | | 2.625 | | 0.875 |
| Case 1-3 | | | | 2.25 | | 0.750 |
| Case1-4 | | | | 1.875 | | 0.625 |
| Case 1-5 | | | | 1.5 | | 0.5 |
| Case 1-6 | | | | 1.125 | | 0.375 |
| Case 1-7 | | | | 0.75 | | 0.25 |
| Case 2-1 | 69.8 | 3.14 | 2 | 6.00 | 0.09 | 0.48 |
| Case 2-2 | | | | 5.25 | | 0.42 |
| Case 2-3 | | | | 4.50 | | 0.36 |
| Case 2-4 | | | | 3.75 | | 0.30 |
| Case 2-5 | | | | 3.00 | | 0.24 |
| Case 2-6 | | | | 2.25 | | 0.18 |
| Case 2-7 | | | | 1.50 | | 0.12 |

Figure 1. A schematic representation of interaction forces between DEM particles *A* and *B*. The gray square represents the mechanical bond between particles *A* and *B*.

Figure 2. The stress-strain curve from a DEM uniaxial tensile simulation with a tensile strength (the maximum tensile stress) of $\sigma_t = 2.7 MPa$.

Figure 3. The stress-strain curve from a DEM uniaxial compressive simulation with a compressive strength (the maximum compressive stress) of $\sigma_c = 9.5 MPa$.

Figure 4. The variation of coefficients *R* and *T* as functions of wave period $\Omega$ for ice thickness of 0.5m, 1m, 2m, and 5m at 100m water depth. (C. Fox and V. A. Squire, J. Geo.Res. Vol. 95 pp. 11636, Copyright 1990, reproduced or modified by permission of AGU).[21]

Figure 5. The geometry of a model sea ice and the DEM particle model.

Figure 6. (color online) The variation of broken bond fraction *α* with dimensionless simulation time *τ* for seven simulation cases in simulation set 1.

Figure 7. (color online) The variation of broken bond fraction *α* with dimensionless simulation time *τ* for seven simulation cases in simulation set 2.

Figure 8a). The variation of maximum broken bond fraction $α_{max}$ with the amplitude $A_1$ of incoming wave for all simulation cases.

Figure 8b). The variation of average fragment length $l$ with the amplitude $A_1$ of incoming wave for all simulation cases.

Figure 9. (color online) The overall response of sea ice for simulation Case 1-1 (an incident wave with $\Omega = 6.28s$ and $A_1 = 3m$). Color represents the stress $\sigma_{xx}$ in $x$ direction normalized by the ice modulus.

Figure 10. (color online) The overall response of sea ice for simulation Case 2-5 (an incident wave with $\Omega = 3.14s$ and $A_1 = 3m$). Color represents the stress $\sigma_{xx}$ in $x$ direction normalized by the ice modulus.

Figure 11. (color online) A snapshot at the end of simulation for Case 1-1 (an incident wave with $\Omega = 6.28s$ and $A_1 = 3m$) showing the spatial distribution of micro-fractures (black arrows) and $\sigma_{xx}$. Color represents the stress $\sigma_{xx}$ in $x$ direction normalized by the ice modulus.

Figure 12. (color online) A snapshot at the end of simulation for Case 1-5 (an incident wave with $\Omega = 6.28s$ and $A_1 = 1.5m$) showing the spatial distribution of micro-fractures (black arrows) and $\sigma_{xx}$. Color represents the stress $\sigma_{xx}$ in $x$ direction normalized by the ice modulus.

Figure 13. (color online) A snapshot at the end of simulation for Case 2-5 (an incident wave with $\Omega = 3.14s$ and $A_1 = 3m$) showing the spatial distribution of micro-fractures (black arrows) and $\sigma_{xx}$. Color represents the stress $\sigma_{xx}$ in $x$ direction normalized by the ice modulus.

Figure 14. (color online) A snapshot at the end of simulation for Case 2-7 (an incident wave with $\Omega = 3.14s$ and $A_1 = 1.5m$) showing the spatial distribution $\sigma_{xx}$ (no micro-fractures observed for this case). Color represents the stress $\sigma_{xx}$ in $x$ direction normalized by the ice modulus.

Fig.1.

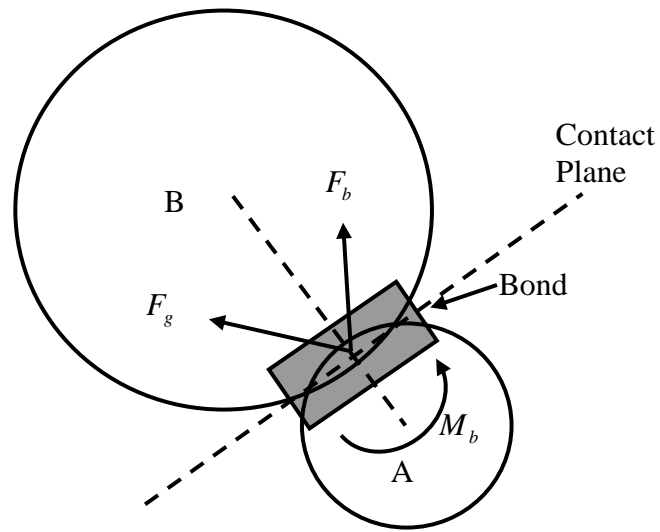

Fig. 2.

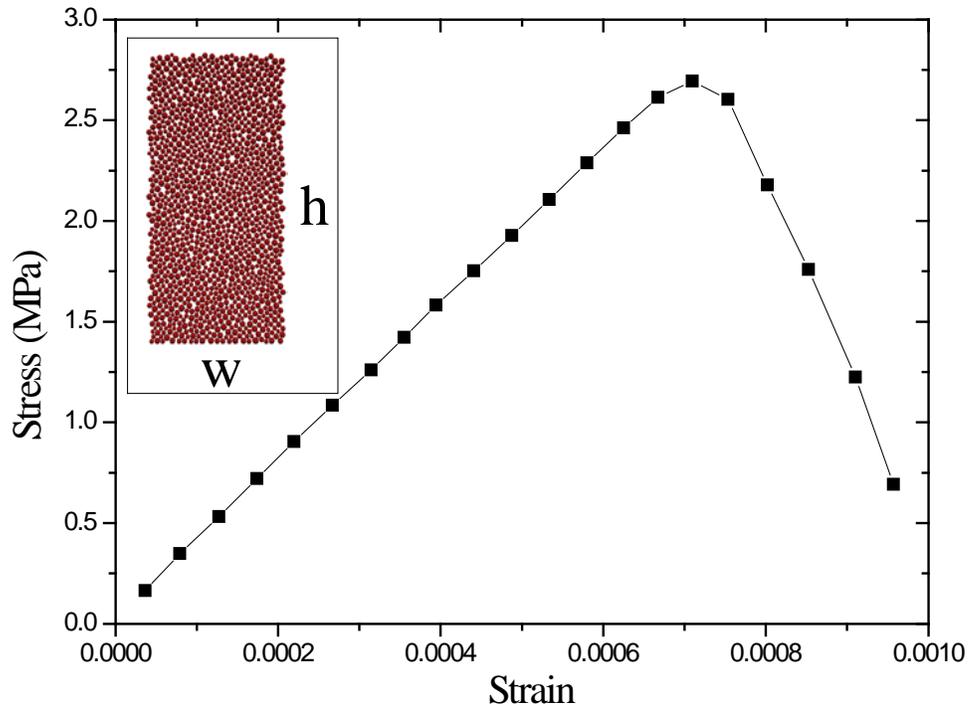

Fig. 3.

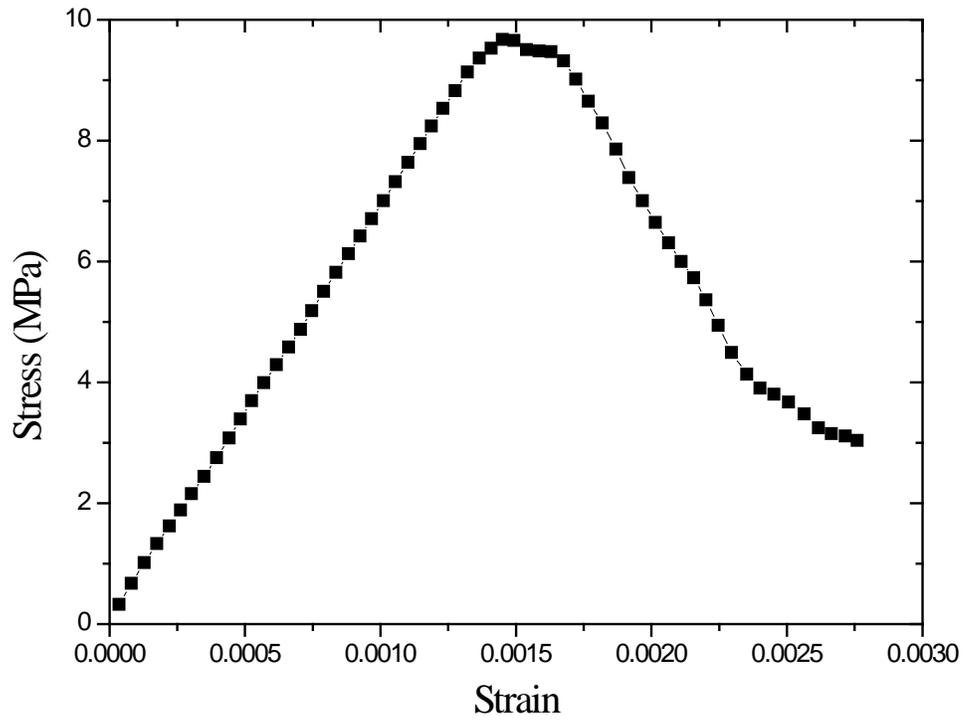

Fig. 4.

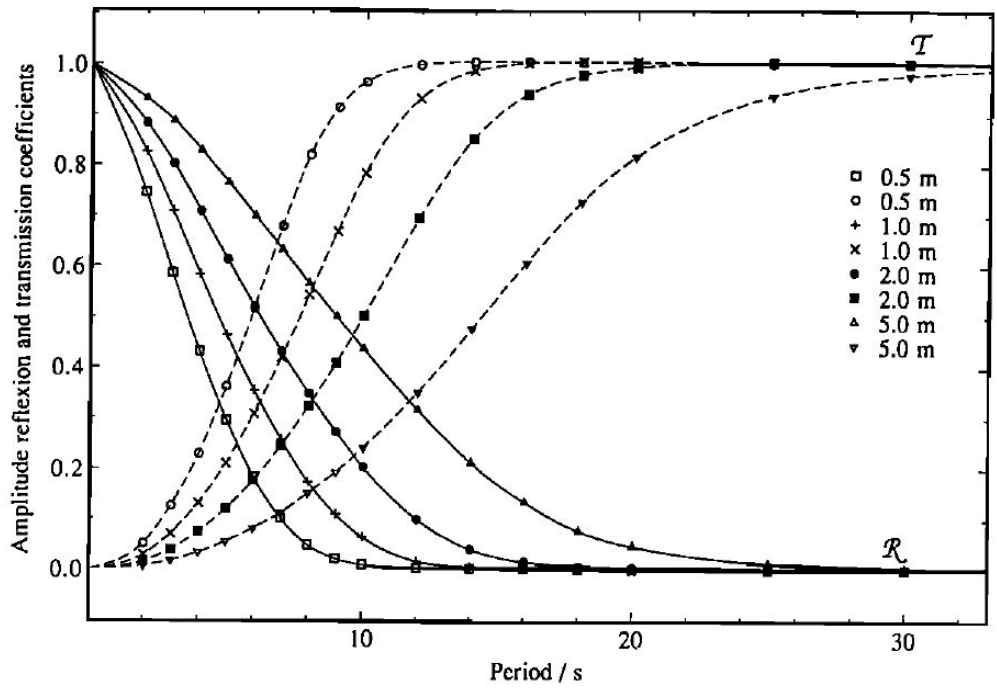

Fig. 5.

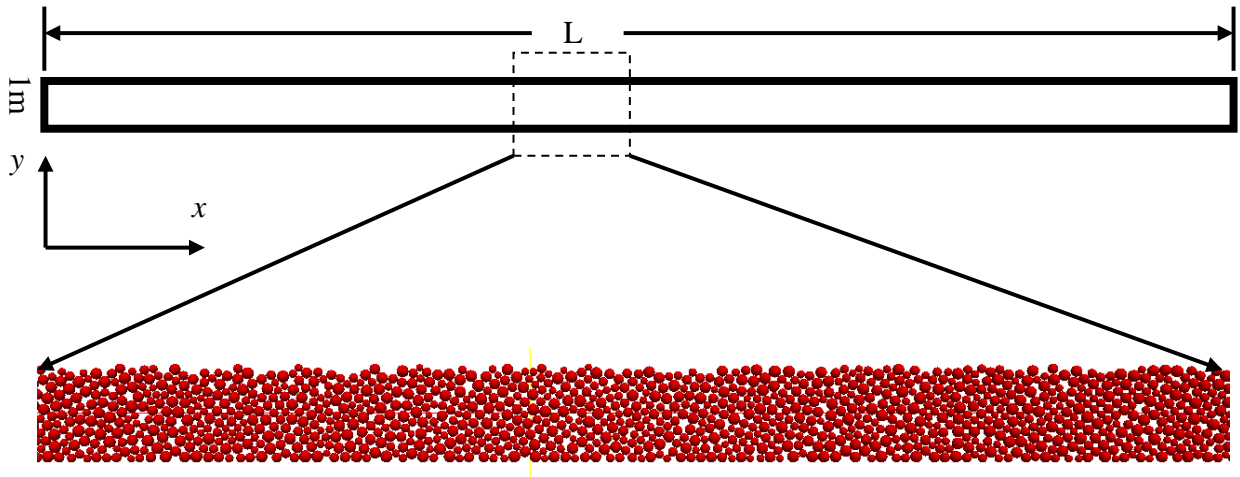

Fig. 6.

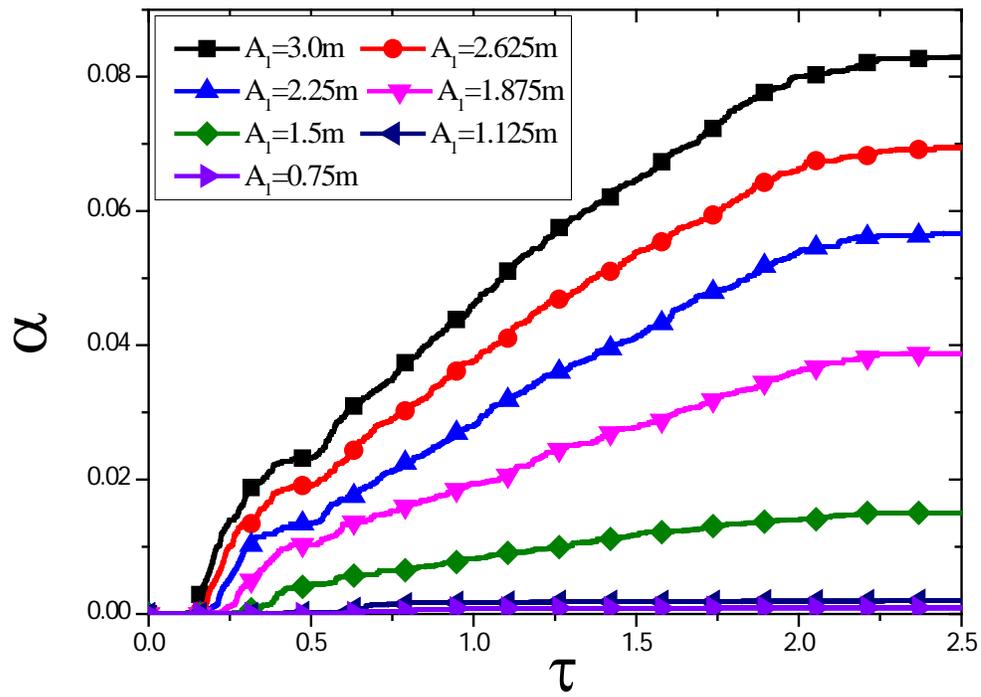

Fig. 7

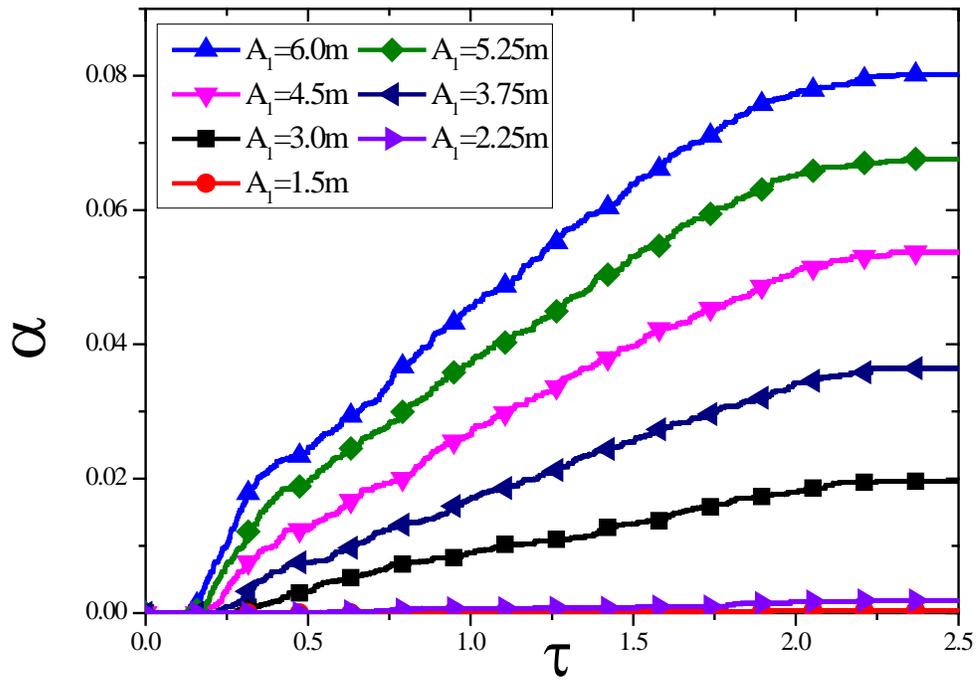

Fig. 8

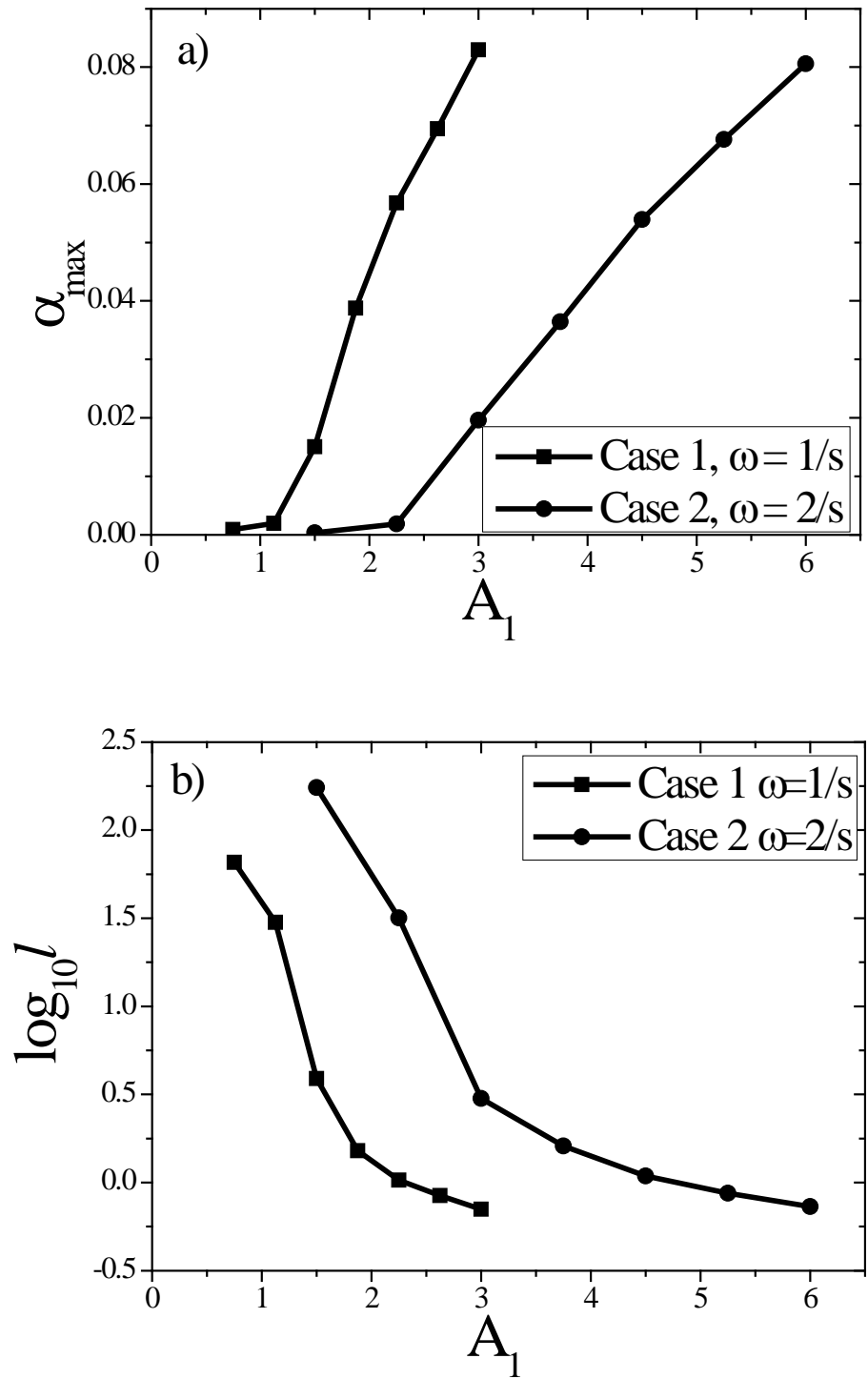

Fig. 9.

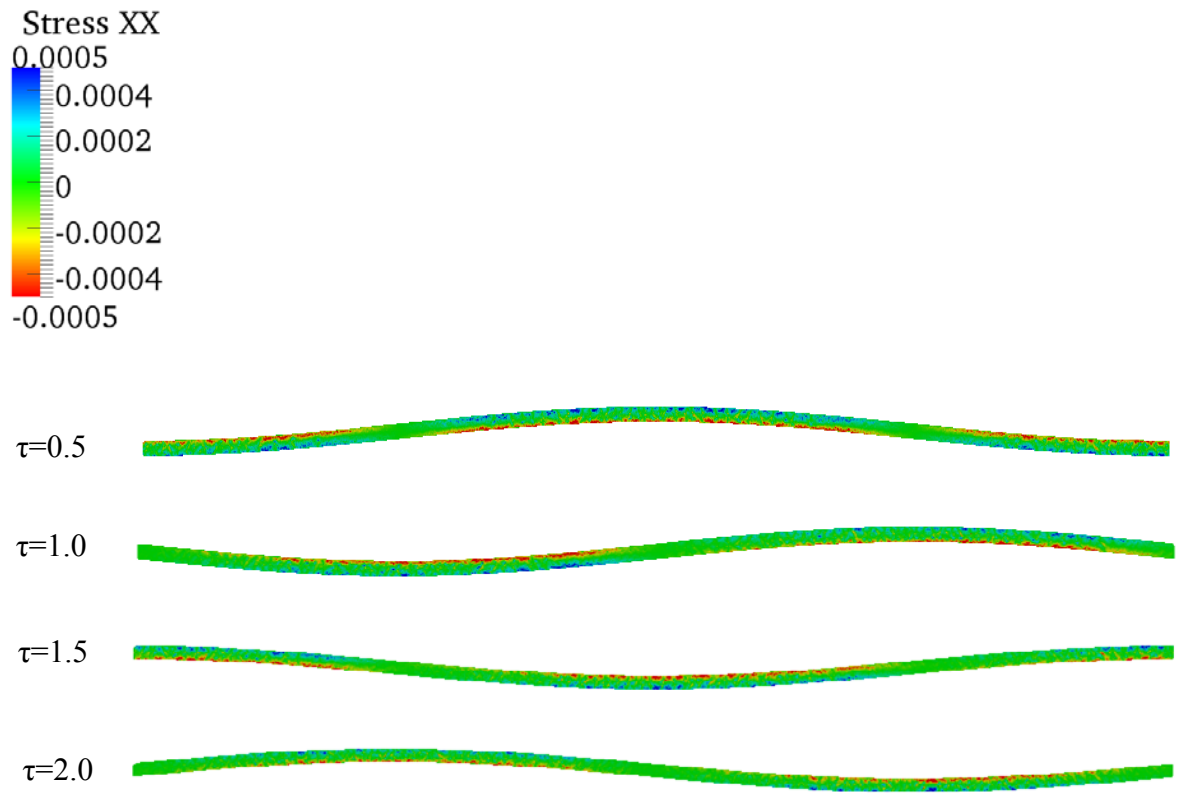

Fig. 10.

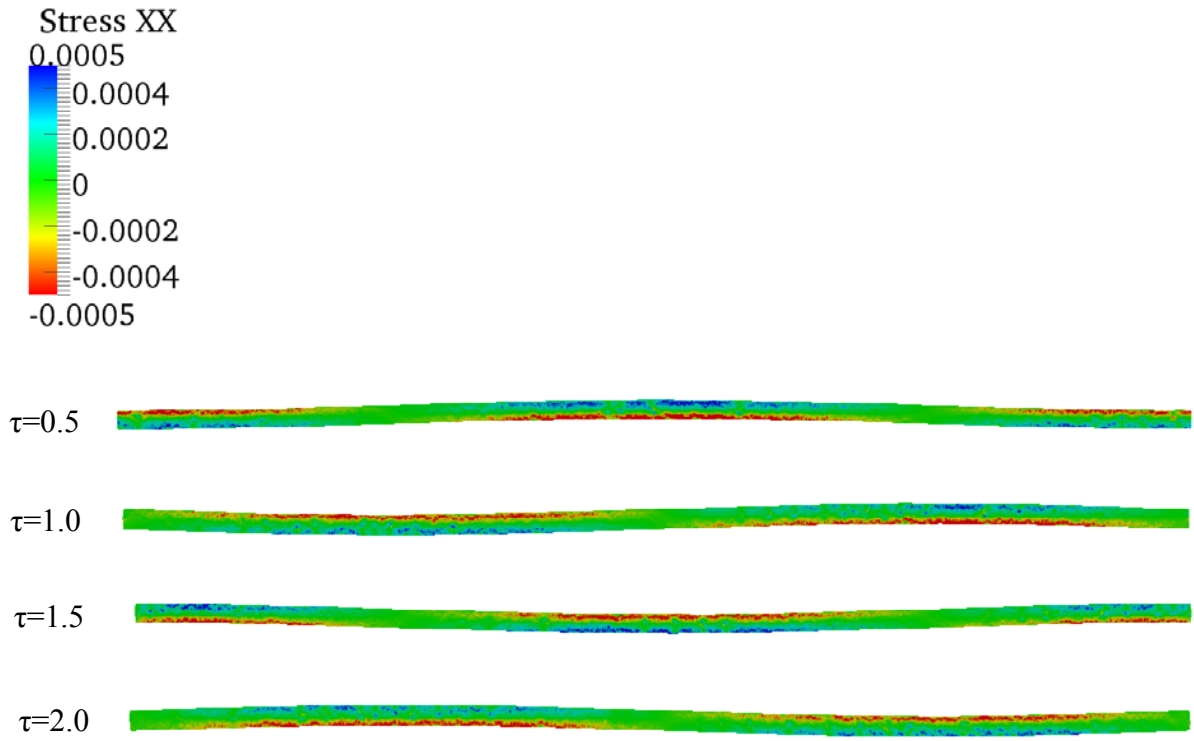

Fig. 11.

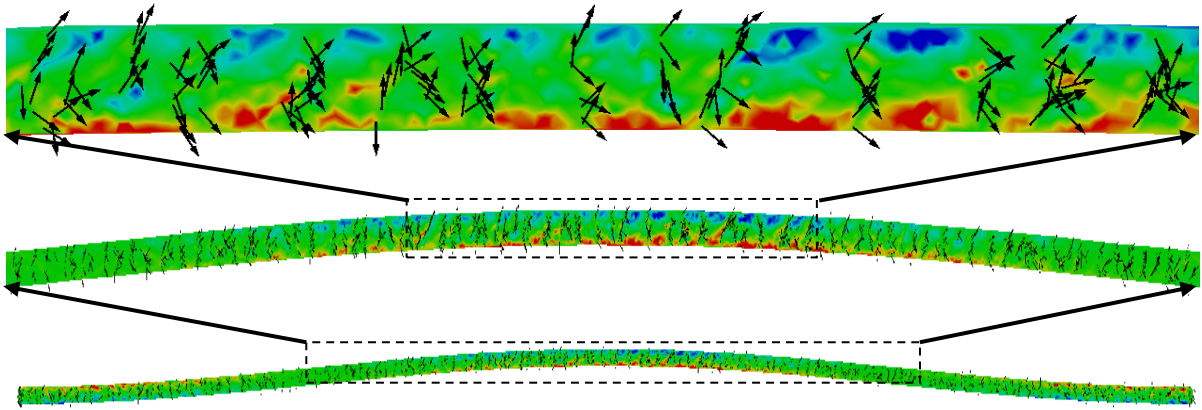

Fig. 12.

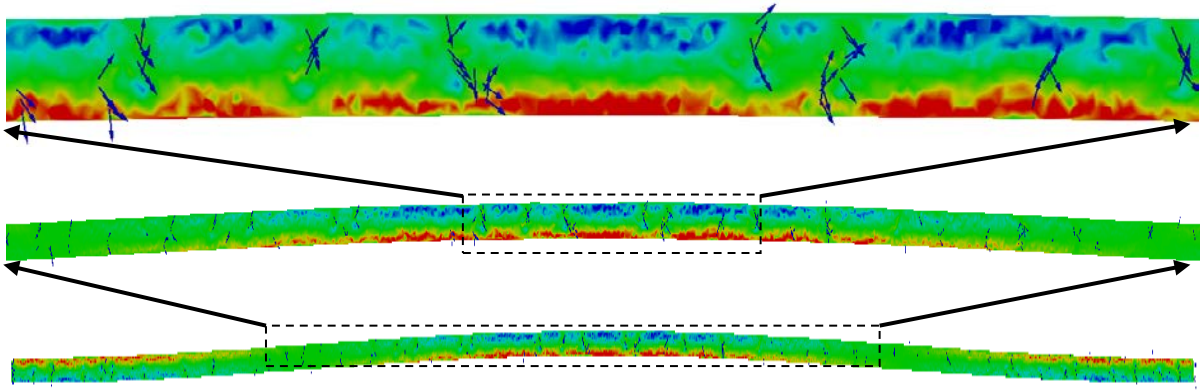

Fig. 13.

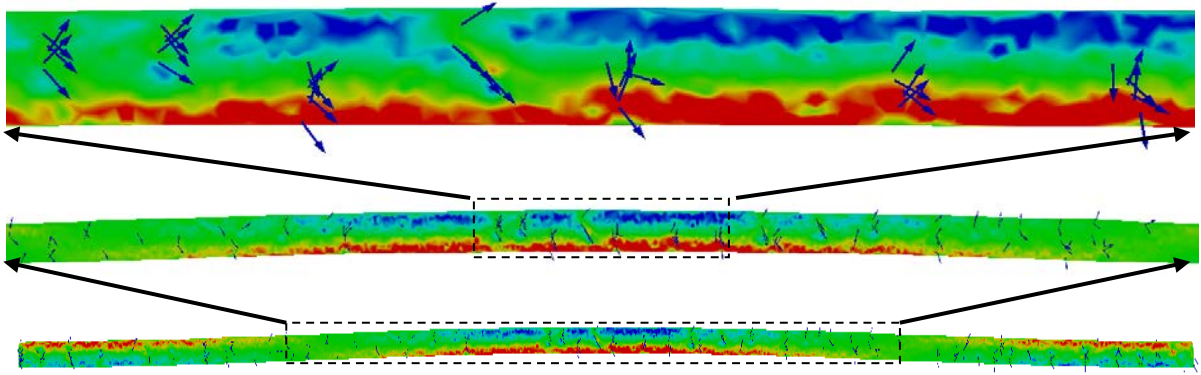

Fig. 14.

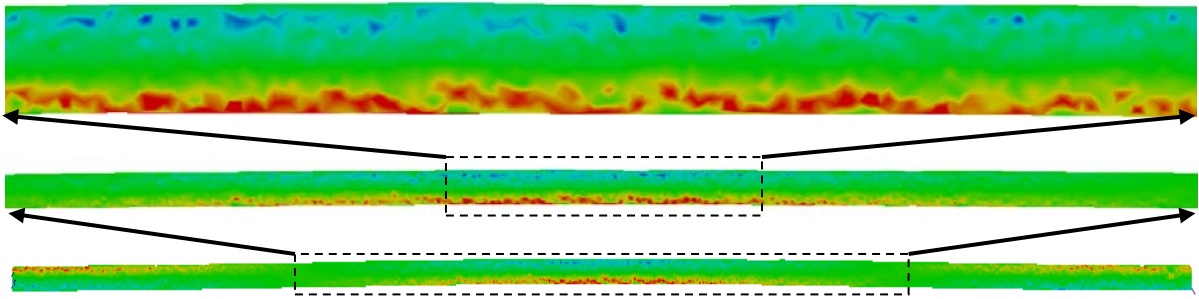